%% file: frw.tex
\documentstyle[psfig,12pt]{article}
\input lmacros

\def\varkappa{\kappa}
\def\fixit#1{}
\def\comment#1{  \begin{raggedright}{\tt [#1]}\end{raggedright}}
\begin{document}
\baselineskip=15.5pt
\pagestyle{plain}
\setcounter{page}{1}

\begin{titlepage}

\begin{flushright}
HUTP-99/A065 \\
hep-th/9912001
\end{flushright}
\vfil

\begin{center}
{\huge AdS/CFT and gravity}
\end{center}

\vfil
\begin{center}
{\large Steven S. Gubser}
\end{center}

$$\seqalign{\span\TL & \span\TT}{
 & Lyman Laboratory of Physics, Harvard University, Cambridge, MA
02138, USA 
  \cr\noalign{\vskip1\jot}
}$$
\vfil

\begin{center}
{\large Abstract}
\end{center}

\noindent 
 The radiation-dominated $k=0$ FRW cosmology emerges as the induced
metric on a codimension one hypersurface of constant extrinsic
curvature in the five-dimensional AdS-Schwarzschild solution.  That we
should get FRW cosmology in this way is an expected result from
AdS/CFT in light of recent comments regarding the coupling of gravity
to ``boundary'' conformal field theories.  I remark on how this
calculation bears on the understanding of Randall and Sundrum's
``alternative to compactification.''  A generalization of the AdS/CFT
prescription for computing Green's functions is suggested, and it is
shown how gravity emerges from it with a strength $G_4 = 2 G_5/L$.
Some upper bounds are set on the radius of curvature $L$ of
$AdS_5$.  One of them comes from estimating the rate of leakage of
visible sector energy into the CFT.  That rate is connected via a
unitarity relation to deviations from Newton's force law at short
distances.  The best bound on $L$ obtained in this paper comes from a
match to the parameters of string theory.  It is $L \lsim 1 \, {\rm
nm}$ if the string scale is $1 \, {\rm GeV}$.  Higher string scales
imply a tighter bound on $L$.

\vfil
\begin{flushleft}
November 1999
\end{flushleft}
\end{titlepage}
\newpage

\section{Introduction}
\label{Introduction}

The AdS/CFT correspondence \cite{juanAdS,gkPol,witHolOne} (see
\cite{MAGOO} for a review) relates a quantum field theory in one
dimension to a theory in one higher dimension that includes gravity.
The primary example is ${\cal N}=4$ super-Yang-Mills theory, which is
related to the dimensional reduction of type~IIB string theory on
$S^5$ to the five-dimensional non-compact geometry $AdS_5$ (anti
de-Sitter space).  The boundary of the Poincar\'e patch of $AdS_5$ is
simply Minkowski space, except that the metric on the boundary is only
specified up to a conformal transformation.  This is OK because ${\cal
N}=4$ super-Yang-Mills theory is conformal, even at the quantum level.
The correspondence is usually studied in a strong coupling region for
the gauge theory, where it is far from classical, but the dual gravity
picture is classical in the sense that curvatures are small on the
Planck and string scales.

In \cite{RSalt} it was proposed that slices of $AdS_5$ could serve as
an alternative to compactification manifolds.  It was shown that when
the near-boundary region of $AdS_5$ is cut away and the bulk spacetime
simply ends on a wall of constant extrinsic curvature (a horosphere of
$AdS_5$ to be precise), there is a normalizable graviton mode which
has zero mass in the four dimensions of the boundary.  The metric of
$AdS_5$ is 
  \eqn{AdSMetric}{
   ds_5^2 = e^{2r/L} (-dt^2 + d\vec{x}^2) + dr^2
  }
 where $\vec{x}$ is an ordinary 3-vector.  This metric is a solution
to $R_{\mu\nu} = -{4 \over L^2} g_{\mu\nu}$.  Hypersurfaces of constant
$r$ are horospheres.  The part of the metric that is cut away in
\cite{RSalt} is $r>r_*$ for some given $r_*$.

In fact the proposal of \cite{RSalt} was to glue two identical copies
of the sliced anti-de Sitter space together along the $3+1$
dimensional boundary.  However the four-dimensional graviton 
is quite a general phenomenon, and there is a
large freedom in what one might have on the other side of the
horospherical boundary of a given copy of $AdS_5$.  An illustration of
this can be found in \cite{HV}, where a single copy of $AdS_5$ is
obtained as part of a type~IIB string compactification on an
orientifold of $T^6$.  The relevant point is that for a vacuum state
of the theory, the extrinsic curvature of the boundary should be
proportional to the induced metric:
  \eqn{ThetaReq}{
   \Theta_{ij} = -{1 \over L} g^{\rm (induced)}_{ij} \,.
  }
 Regarding the boundary as a positive tension $3+1$-dimensional brane
separating two sliced copies of $AdS_5$, this amounts to the statement
that the stress energy of the brane should respect $3+1$-dimensional
Poincar\'e invariance.  The constant of proportionality, $1/L$, is
required so that there is a balance between the brane tension and the
bulk cosmological constant.  More generally, a codimension one
boundary of a five-dimensional space with no excitations on it should
have the same property that $\Theta_{ij} = -{1 \over L} g^{\rm
(induced)}_{ij}$.

The authors of \cite{RSalt} termed their construction an ``alternative to
compactification,'' which seemed appropriate because one can travel an
infinite spatial distance into the five-dimensional bulk.  Efforts such as
\cite{HV} to realize the construction in string theory, using coincident
D3-branes to produce the $AdS_5$ background, represent strong-coupling
extrapolations of perturbative string compactifications, where the massless
four-dimensional graviton is the usual zero-mode of the spin 2 closed
string state.  In such a picture, the Kaluza-Klein gravitons of
\cite{RSalt} are interpreted as the strong-coupling description of open
string excitations on the D3-branes, in line with the extensive literature
on absorption by D3-branes \cite{IgorAbsorb,gktAbsorb,gkSchwing}.

It was suggested by J.~Maldacena \cite{juanPrivate} that this ``alternative
to compactification'' should properly be viewed in light of the AdS/CFT
correspondence as a coupling of gravity to whatever strongly coupled
conformal theory the $AdS_5$ geometry is dual to.  This view was also taken
by H.~Verlinde in \cite{HV,VerlindeTalk}.  A convincing statement of the
case was made by E.~Witten \cite{WittenComment} in response to
\cite{SundrumTalk,GiddingsTalk}.

From now on I will restrict myself to a minimal scenario where a single
copy of $AdS_5$ is cut off by an end-of-the-universe brane.  Such objects
are well known in type~I$'$ string theory \cite{PolchinskiWitten} and in
Horava-Witten theory \cite{HoravaWitten}, so there is no problem of
principle in having a true end of the universe.  However my comments
basically apply to any compactification geometry which involves the
near-horizon part of $AdS_5$.

The idea that the scenario of \cite{RSalt} is best viewed in the
context of AdS/CFT has not been universally embraced, perhaps partly
because it is hard to see what to do with it.  (That difficulty is not
usually regarded as fatal, but it does seem to have held up progress
on the current issue).  The goal of section~\ref{FRW} is to make the
idea seem more definite by using it to derive the radiation-dominated
Friedmann-Robertson-Walker (FRW) cosmology.  The approach is to change
the bulk spacetime from $AdS_5$ to AdS-Schwarzschild, but not to
excite any matter on the cutoff brane.  The Hawking temperature of the
AdS-Schwarzschild geometry, measured with respect to time on the
cutoff brane, can be interpreted as the temperature of the CFT which
the bulk spacetime is dual to.  Readers determined to understand the
construction from a brane-world perspective may find it most useful to
think of the bulk as the background of near-extremal D3-branes.
However, AdS/CFT allows us to reinterpret the entire AdS-Schwarzschild
geometry as a manifestation of the dynamics of a four-dimensional
conformal field theory at finite temperature.  With the CFT
interpretation in mind, it is easier to understand why the
radiation-dominated FRW geometry emerges: all CFT's have the same
equation of state up to numerical factors, so the FRW equations take
the same form as they do in the radiation-dominated era of our
universe.

The literature on brane-world cosmology is large and somewhat
scattered; indeed I became aware of some recent overlapping papers
only after this work was completed.  Early work on supergravity domain
walls in $AdS_4$ has been reviewed in \cite{Cvetic:1993xe}.
Cosmological solutions to Horava-Witten theory compactified on a
Calabi-Yau manifold were investigated in \cite{Lukas:1998qs}; see also
\cite{Chamblin:1999ya} for subsequent development, and
\cite{Chamblin:1999ea} for an analogous treatment of type~I$'$.  The
works
\cite{KrausCosm,Mukohyama:1999qx,Binetruy:1999hy,Kehagias:1999vr}
include constructions equivalent to equations
\eno{Foliate}-\eno{TauForm}, although the CFT interpretation was not
offered.  Reference \cite{Shiromizu:1999wj} includes some formal developments
and presents equations equivalent to \eno{EinsteinBold}.  Discussions
of cosmological constraints in \cite{Binetruy:1999hy,Flanagan:1999cu}
have some overlap with section~\ref{Estimate}.  And there are a number
of papers \cite{Nihei:1999mt,Kaloper:1999sm,DFGK,Csaki:1999mp,Vollick}
that use similar brane-world techniques.  A more extensive list of
references on brane-world cosmology can be found in
\cite{Csaki:1999mp}.  The whole approach is rather different from the
older string cosmology literature; see \cite{Brustein} for references.

Section~\ref{General} consists of some remarks on the general
framework of AdS/CFT with a cutoff brane.  A generalization of the
prescription \cite{gkPol,witHolOne} for computing Green's functions is
suggested at the level of effective field theory.  The
four-dimensional Einstein action can be derived in this formalism,
with the result $G_4 = 2 G_5/L$.  This relation obtains regardless of
the location of the cutoff brane in $AdS_5$.  In standard AdS/CFT,
where there is no cutoff brane and hence no normalizable graviton, the
terms responsible for the Einstein action were removed using local
counter-terms \cite{hs}.  Corrections to Newton's force law are
discussed from the CFT perspective, 
but I avoid presenting details since the idea of the
calculation is not original to me.

Matter on the cutoff brane is incorporated naturally in the formalism.
Although I do not propose a definite model, the idea is to have
visible sector matter on the cutoff brane, somewhat as in certain
heterotic M-theory models \cite{OvrutLecture}.  Excitations of that
matter would have to dominate over the solution in section~\ref{FRW}
at least for $z \lsim 10^{10}$ in order for the cosmology to be realistic.

In section~\ref{Estimate}, I make some rough numerical estimates.  One
is to check the cosmological effects of visible sector matter losing
its energy to the CFT.  In AdS language this corresponds to absorption
of bulk gravitons by the horizon of the Poincar\'e patch, as in
\cite{IgorAbsorb,gktAbsorb,gkSchwing}.  The rate of energy loss is
directly related to the deviations from Newton's force law.  Another
is to estimate parameters of string theory that would permit the
deviations from Newton's force law to be observed experimentally,
assuming that $AdS_5$ emerges from a string theory construction.  To
obtain deviations at the scale of even a nanometer (still three orders
of magnitude below the sensitivity of proposed experiments) an
extremely low string scale is required---approximately $1 \, {\rm
GeV}$.  I make some speculative remarks regarding low string scales at the
end of section~\ref{Estimate}.

Throughout the paper, $\mu$ will denote a five-dimensional bulk
spacetime index and $i$ is a four-dimensional index.  In cases where
precision is required, I will denote five-dimensional coordinates as
$x^\mu = (t,\vec{x},r)$ and four-dimensional coordinates as $\xi^i =
(\tau,\vec{\xi})$.  Throughout the paper, $\ell_{\rm Pl}$ will denote
the four-dimensional Planck length: $\ell_{\rm Pl} = \sqrt{G_4}
\approx 1.6 \times 10^{-33} \, {\rm cm}$ in units where $\hbar = c =
1$.

\section{A cosmological solution}
\label{FRW}

Let us start purely from a four-dimensional point of view, and turn on
a finite temperature for the conformal field theory which is small in
Planck units.  If we calculate the corresponding energy density
$\rho$, use the trivial equation of state $p = \rho/3$, and apply the
standard equation
  \eqn{aEOM}{
   \left( \dot{a} \over a \right)^2 = {8\pi G_4 \over 3} \rho \,,
  }
 what must result is the standard radiation-dominated FRW cosmology,
  \eqn{FRWMetric}{
   ds^2 = -d\tau^2 + a(\tau)^2 d\vec{x}^2 \,,
  }
 where $a(\tau)^2$ is linear in $\tau$.\footnote{There is probably no
obstacle in principle to extending the calculation to positive spatial
curvature.  However for negative spatial curvature the
pathology observed in \cite{WittenYau} might emerge.}  I will always
use $\tau$ for four-dimensional cosmological time; $t$ will be
reserved for the Poincar\'e time in $AdS_5$.  The only difficulty is
finding the constant of proportionality in $\rho \propto T^4$: if the
conformal field theory is interacting, it could be a non-trivial
exercise even in the limit $T \ll 1/\ell_{\rm Pl}$ where gravity loops
shouldn't matter.  However if the theory in question has an AdS dual
where the supergravity approximation is good,
then the study of black holes in AdS guarantees the relation
  \eqn{RhoRel}{
   \rho = {3 \pi^2 \over 2} c T^4 \,,
  }
 where $c$ is the coefficient for the trace anomaly in a normalization
where $c = (N^2-1)/4$ for ${\cal N}=4$ $SU(N)$ super-Yang-Mills
theory.\footnote{The calculation of \cite{hs} shows that in the limit
where classical gravity is applicable to the AdS black holes there is
in fact only one independent coefficient in the trace anomaly.  This
and \RhoRel\ are non-trivial constraints on theories which can have
AdS duals.}  In this normalization, a single Abelian photon has $c =
1/10$.  There is a standard relation in AdS/CFT \cite{hs}
  \eqn{cStandard}{
   c = {\pi \over 8} {L^3 \over G_5} \,.
  }
 It should be remarked that for ${\cal N}=4$ gauge theory at weak
coupling, \RhoRel\ becomes $\rho = 2\pi^2 c T^4$.  This is the $4/3$
problem, first noted in \cite{gkPeet,StromingerUnp}, and now
understood as being a result of strong interactions.

So far we have employed the AdS/CFT correspondence merely as a tool for
determining a detail of the strong-coupling thermodynamics.  However the
calculation can be done entirely on the AdS side if we take seriously the
idea that the cutoff brane is no more nor less than a coupling of gravity
to the conformal field theory.  It seems inevitable from a string theory
perspective that the gravity would be quantized, and that the detailed
``structure'' of the cutoff brane encodes the details of the quantum
gravity; but by taking $T \ll 1/\ell_{\rm Pl}$ we should be able to ignore
this issue.  The cutoff brane, or ``Planck brane,'' controls gravity, while
the bulk of $AdS_5$ controls the conformal field theory.  By assumption,
the Planck brane is not appreciably influenced by finite temperature, but
the conformal field theory is; so we should retain \ThetaReq, but change
the bulk background from $AdS_5$ to AdS-Schwarzschild.  The metric of
AdS-Schwarzschild is
  \eqn{AdSSch}{\eqalign{
   ds_5^2 &= e^{2r/L} \left( -h(r) dt^2 + d\vec{x}^2 \right) +
    {dr^2 \over h(r)}  \cr
   h(r) &= 1 - b^4 e^{-4r/L}  \cr
   b &= \pi L T_0 \,.
  }}
 Here $T_0$ is the Hawking temperature associated with the time
coordinate $t$.  It is a constant parameter of the
AdS-Schwarzschild solution.  The calculation will deal only with
the coordinate patch covered by $(t,\vec{x},r)$.

Given an orientable surface with unit normal $n_\mu$ (which specifies a
notion of outside and inside by the direction in which it points), the
extrinsic curvature can be defined as 
$\Theta_{\mu\nu} = -(\delta_\mu^\lambda -
n_\mu n^\lambda) \nabla_\lambda n_\nu$.  
We will follow \cite{BK} in taking $n_\mu$ to
be the outward unit normal, which points toward the true boundary of
$AdS_5$.  There is a set of solutions to \ThetaReq\ which form a
foliation of the coordinate patch in question:
  \eqn{Foliate}{
   {t \over L} = {e^{r/L} \over b^2} + 
    {1 \over 4b} \sum_{k=1}^4 i^k \log (1-i^k b e^{-r/L}) + 
    {t_0 \over L} \,,
  }
 where $t_0$ is a constant of integration specifying a particular
leaf.  All leaves have the same induced metric since they are related
by translation in $t$.  Using \ThetaReq\ with the same constant of
proportionality, $1/L$, avoids a four-dimensional cosmological
constant---more about this later.  It proves most convenient to
parametrize a particular leaf by $(r_*,\vec{x})$, where $r = r_*$ is a
solution of \Foliate\ for $r$ in terms of $t$.  Then the induced
metric is
  \eqn{InducedG}{
   ds_{\rm (induced)}^2 = 
    -{e^{4r_*/L} \over b^4} dr_*^2 + e^{2r_*/L} d\vec{x}^2 \,.
  }
If we define
  \eqn{TauDef}{
   \tau = L {e^{2r_*/L} \over 2 b^2}
  }
 then the metric \InducedG\ assumes the standard FRW form:
  \eqn{TauForm}{\eqalign{
   ds_{\rm (induced)}^2 &= -d\tau^2 + a(\tau)^2 d\vec{x}^2  \cr
   a(\tau) &= b \sqrt{2 \tau \over L} \,.
  }}
 Thus we do indeed observe the linear $a(\tau)^2$ that we expected.
This behavior is strictly a consequence of conformal invariance: any
conformal field theory provides a source term for Einstein's equations
just like a bunch of massless photons.  

We can be a little more quantitative and rederive the coefficient in
\RhoRel\ from the new perspective.  In the late time limit, we can use
the relation
  \eqn{GFiveFour}{
   {1 \over G_4} = {1 \over G_5} 
    \int_{-\infty}^{r_*} dr \, e^{2 (r-r_*)/L}
    = {L \over 2 G_5} \,.
  }
 Actually this relation comes from a Kaluza-Klein reduction of
five-dimensional gravity to four for a horospherical Planck brane in
pure $AdS_5$.  It should be OK to leading order for a brane in an
asymptotically $AdS_5$ region of bulk spacetime, provided the brane is
only slightly curved on the scale $L$.  Such a brane is locally like a
horosphere of $AdS_5$.  It is perhaps more common in the literature to
see a modification of \GFiveFour\ that arises from removing the factor of 
$e^{-2 r_*/L}$ from inside the integral.
The discrepancy is merely due to the fact that our four-dimensional
metric is precisely the induced metric, whereas more commonly the
four-dimensional metric on a horosphere is taken to be $e^{-2 r_*/L}
ds_{\rm (induced)}^2$.  The form of \GFiveFour\ is forced on us by the
choice $ds_4^2 = ds_{\rm (induced)}^2$, which does seem the natural
one in the present context.

Combining the relation
  \eqn{LateTimes}{
   {1 \over 4\tau^2} = \left( {\dot{a} \over a} \right)^2 = 
    {8\pi G_4 \over 3} \rho
  }
 with \GFiveFour\ and \cStandard\ leads to 
  \eqn{RhoLate}{
   \rho = {3 b^4 \over 16 \pi G_5 L e^{4 r_*/L}} = 
    {3\pi^2 \over 2} c (e^{-r_*/L} T_0)^4 \,.
  }
 Now, the temperature $T_0$ measured with respect to the time $t$ is
not the same as the temperature $T$ measured with respect to the time
$\tau$; rather, 
  \eqn{TTRel}{
   T = {dt \over d\tau} T_0 = e^{-r_*/L} T_0 \,,
  }
 where we have used the relation $t = \sqrt{2L\tau}/b$.  So \RhoLate\
is indeed identical to \RhoRel, coefficient and all.

The foregoing calculation is more than a formal manipulation: it is an
illustration that {\it string theory on $AdS_5$ is
identical to a 3+1-dimensional conformal field theory.}  
We wanted our cosmology to be driven by the
conformal field theory dual to the bulk AdS geometry rather than by
anything on the Planck brane.  So we left the Planck brane in its
ground state and made the bulk AdS geometry thermal by adding a black
hole horizon.

It is worth remarking that no stabilization mechanism was employed
because none was needed.  From a brane-world point of view, the worst
has already happened: the negative tension brane of \cite{RShierarchy}
has retreated to infinity,\footnote{The reality of such a negative
tension object is something I am only provisionally willing to allow
for the sake of argument, since I am aware of no fully satisfactory
string theory construction of it as yet in an AdS background.} and the
delicate near-horizon cusp has been cut off by a finite temperature
horizon.  The effect of that horizon is most transparent when viewed
in light of the AdS/CFT correspondence: it means that the
$3+1$-dimensional conformal field theory is at finite temperature.

In a generic bulk geometry, the retreating Planck brane would cause
the four-dimensional Newton constant to change.  In this regard, an
asymptotically AdS space is very special: provided we use the induced
metric on the Planck brane (rather than some warping of it) as the
four-dimensional metric, \GFiveFour\ will hold asymptotically
when the Planck brane is moving in the asymptotically AdS region with
curvatures which are small compared to $L$.

That $ds_{\rm (induced)}^2$ turned out to be {\it exactly} the
radiation-dominated FRW metric should excite some suspicion.
Mightn't there be quantum gravity effects at
sufficiently early times which modify the picture?  We derived the
agreement between \RhoRel\ and \RhoLate\ using a late-time relation,
\GFiveFour.  It seems like a massive conspiracy that the physics of
early times would arrange for the cosmology to remain exactly
radiation-dominated FRW.  There are limits to what we can assert about 
physics at early times without specifying the nature of the Planck brane.
It seems inevitable however that \ThetaReq\ will receive corrections at
higher orders in derivatives.

Is this real cosmology?  Not as it stands: nucleosynthesis would be
dramatically spoiled if the ``hidden CFT'' that $AdS_5$ represents had
any sizable effect on the radiation-dominated era of our universe.
However it is straightforward to extend the discussion by adding
matter to the brane, and its stress tensor, $T_{ij}^{\rm (matter)}$,
could take over from $T_{ij}^{\rm (CFT)}$ at late times.  (If all we're
worried about is nucleosynthesis, then late times means $z \lsim
10^{10}$).  The AdS/CFT equation relevant to such a scenario is
  \eqn{EinsteinBold}{\eqalign{
   &G_{ij}^{\rm (induced)} - 8 \pi G_4 T_{ij}^{\rm (CFT)} - 
    8 \pi G_4 T_{ij}^{\rm (matter)} =  \cr
   &\qquad\quad
    -{2 \over L} \left[ \Theta_{ij} - \left( \Theta + 
     {3 \over L} \right) g_{ij}^{\rm (induced)} \right] - 
     8 \pi G_4 T_{ij}^{\rm (matter)} \,.
  }}
 This equation is a rearrangement of a formula obtained in \cite{BK}
in the course of deriving quasi-local stress-energy tensors for
AdS/CFT in various dimensions.\footnote{Only I have changed the sign
of the Einstein tensor.  This is necessary because of a difference
in sign conventions.  The derivation in section~\ref{General} will
serve as a check that the signs in \EinsteinBold\ are consistent.}  I
have used \GFiveFour\ to define $G_4$.  It is necessary to check that
visible matter does not lose energy to the CFT fast enough to spoil
the cosmology.  Since the CFT's couplings are essentially of
gravitational origin, this is perhaps plausible.  An estimate will be
presented in section~\ref{Estimate}.

Actually, \EinsteinBold\ is only an approximate statement of translation,
via AdS/CFT, from purely four-dimensional quantities to quantities which
constrain how the Planck brane sits in the five-dimensional spacetime.
Equations with physical meaning arise from setting either side equal to
zero.  In this section we considered a case where $T_{ij}^{\rm (brane)} =
0$; then setting the right hand side equal to zero amounts to requiring
\ThetaReq.  Solving \ThetaReq\ gave us back radiation-dominated FRW
cosmology, which perhaps sounded surprising; but the identity
\EinsteinBold\ makes it inevitable, because if the right hand side
vanishes, so must the left hand side.

Suppose now we knew all about the matter on the brane, and discovered
that it generated a positive cosmological constant: $-8 \pi G_4
T_{ij}^{\rm (matter)} = \Lambda g_{ij}^{\rm (induced)}$ with $\Lambda
> 0$.  Assuming the AdS part to be at zero temperature, we would then
reduce \EinsteinBold\ to
  \eqn{EBSimp}{
   G_{ij}^{\rm (induced)} + \Lambda g_{ij}^{\rm (induced)} = 
    -{2 \over L} \left[ \Theta_{ij} - \left( \Theta + 
     {3 \over L} + {L^2 \Lambda \over 2} \right) 
     g_{ij}^{\rm (induced)} \right] \,.
  }
 It is straightforward to show that setting the right hand side equal
to zero leads to a hypersurface in $AdS_5$ whose induced metric is
$dS_4$.  Approximately this calculation has appeared elsewhere in the
literature, for instance
\cite{Nihei:1999mt,Kaloper:1999sm,DFGK,Vollick}.  A direct analog in
lower dimensions was treated in \cite{Cvetic:1993xe}, where a fairly
general discussion of induced metrics on codimension one domain walls
in $AdS_4$ was also given.

\section{The general framework}
\label{General}

The equation \EinsteinBold\ is an approximate first variation of a
more general relation, which is the natural extension of the
prescriptions of \cite{gkPol,witHolOne}:
  \eqn{DarkSide}{\eqalign{
   S_{\rm eff}[\gamma_{ij},\psi] &=
     S_{\rm 4d\,gravity}[\gamma_{ij}] + 
     S_{\rm 4d\,matter}[\gamma_{ij},\psi] +
     W_{\rm CFT}[\gamma_{ij}]  \cr\noalign{\vskip1.5\jot}
    &= \mathop{\rm extremum}_{g_{ij}^{\rm (induced)} = \gamma_{ij}} \left(
       S_{\rm bulk}[g_{\mu\nu}] + S_{\rm brane}[g_{ij}^{\rm (induced)},\psi]
       \right) \,.
  }}
 The metric
$\gamma_{ij}$ is the metric on the Planck brane, and $\psi$ are the
extra matter fields which live on the Planck brane.  
The first equation indicates a natural way of splitting up the
four-dimensional effective action into four-dimensional gravity, the
CFT, and the four-dimensional matter which comes from excitations on
the Planck brane.  The second equation is the actual statement of
AdS/CFT, which in this case includes a ``brane reduction'' of
five-dimensional gravity to four dimensions, as envisaged in
\cite{RShierarchy,RSalt}.  $W_{\rm CFT}$ is the generating functional
of connected Green's functions of the conformal field theory, with a cutoff
imposed at some energy scale $\Lambda$.  See \cite{SusskindWitten} for
an early discussion of cutoffs in AdS/CFT.

One way to define the cutoff $\Lambda$ is as the energy of a fundamental
string stretched from the Planck brane all the way to the horizon of
$AdS_5$.  This gives $\Lambda \sim L/\alpha'$ if we measure energies with
respect to a time $\tau$ on the Planck brane such that $\gamma_{\tau\tau} =
-1$.  There can however be ambiguities in normalizing $\Lambda$, depending
on the physical question one is asking, as explained in
\cite{PeetPolchinski}.  If $AdS_5$ is generated as the background of many
coincident D3-branes, then we can imagine peeling one of them off and
bringing it close to the Planck brane.  A fundamental string stretched from
this D3-brane back to the main cluster has the interpretation of a massive
W-boson.  In the supergravity description, this fundamental string
stretches to the horizon of $AdS_5$.  Thus the relation $\Lambda \sim
L/\alpha'$ has a simple motivation in terms of a BPS quantity, namely the
mass for the heaviest W-boson which can be included in the effective theory
by Higgsing the CFT.  A comprehensive discussion of interactions might
require a more precise specification of how a geometric cutoff in $AdS_5$
translates into a cutoff in the four-dimensional theory.

The extremum on the right hand side of \DarkSide\ 
is taken subject to the boundary condition that
the metric induced from $g_{\mu\nu}$ on the cutoff brane is
$\gamma_{ij}$.  It is the saddle-point approximation to quantum
gravity in the bulk.  If we wanted to do quantum gravity in some more
complete way (i.e.~string theory), we would make the replacement 
  \eqn{Quantize}{
   {\rm extremum} \, S \to {1 \over i} \log \int [{\cal D} g] e^{i S} \,,
  }
 where $\int [{\cal D} g]$ represents path integration.  (Path integration
in the sense of \Quantize\ would amount to closed string field theory---a
subject where our understanding is incomplete.  We might however imagine some
other way of improving the saddle point approximation).  In a real
string theory model, there would probably be many more bulk fields besides
the metric $g_{\mu\nu}$ that $S_{\rm bulk}$ would depend on, and they would
also have to have their boundary values specified in the extremum (or path
integral).  $W_{\rm CFT}$ would depend on those boundary values, and there
would also be new terms added to $S_{\rm 4d\,gravity}$ for the dynamics of
the zero modes of the extra fields.  How much of a problem all this extra
junk is depends on the couplings to the standard model fields.  The
optimistic view is that such couplings are about as important in particle
physics contexts as the coupling of electrons and quarks to gravity.  The
zero modes of extra bulk fields would modify long-distance four-dimensional
gravity if they remained massless, but any sort of confinement or mass
generation mechanism could prevent this problem.

Like all of AdS/CFT, \DarkSide\ is a claim to be substantiated rather
than an assumption.  However, it is difficult to give a complete proof
because
$W_{\rm CFT}$ is a complicated non-local functional of $\gamma_{ij}$
whose exact form is independently accessible only through
a strong coupling QFT computation.  If one takes the boundary to be
the true boundary of $AdS_5$, the evidence is compelling \cite{MAGOO}
that the extremum on the right hand side of \DarkSide\ does indeed
lead to the generating functional of connected Green's functions for a
CFT.\footnote{A conformal transformation on $\gamma_{ij}$ is needed 
as the cutoff is removed to keep $\gamma_{ij}$ finite.  However, 
$W_{\rm CFT}$ without a cutoff only depends on the conformal class of
$\gamma_{ij}$.}  Through the UV-IR relation we understand that cutting off a
portion of $AdS_5$ should change physics in the ultraviolet only.
Thus \DarkSide\ is true insofar as it is well-defined (that is, on the
level of an effective field theory on energy scales much lower than the
cutoff $\Lambda$) provided we can show that $S_{\rm
4d\,gravity} + S_{\rm 4d\,matter}$ emerges from the extremum on the
right hand side.  That is what I will actually demonstrate concretely.
In the process I will derive \GFiveFour\ in a general setting, and
also check that the sign that seemed worrisome in \EinsteinBold\ is
OK.

The proof is piggy-backed on the calculations of \cite{hs}.  In order
to keep the presentation self-contained, I will recapitulate parts of
that work.  The setting is a foliation
of a five-dimensional Einstein manifold ${\cal M}$ (for instance,
$AdS_5$ or AdS-Schwarzschild), whose boundary has a metric in the
conformal class of a specified metric $\bar{g}_{(0)}$, and whose metric can
be written in the form \cite{FeffGraham}
  \eqn{FeffMetric}{
   ds_5^2 = g_{\mu\nu} dx^\mu dx^\nu = {1 \over 4 \rho^2} d\rho^2 + 
    {1 \over \rho} \bar{g}_{ij} d\xi^i d\xi^j \,.
  }
 In \FeffMetric\ and the following equations, $L$ has been set to
$1$.  The metric $\bar{g}_{ij}$ can depend on $\rho$, but according to
\cite{hs,FeffGraham} it has an expansion
  \eqn{gExpand}{
   \bar{g} = \bar{g}_{(0)} + \rho \bar{g}_{(2)} + 
    \rho^2 \log\rho \, \bar{h}_{(4)} + \rho^2 \bar{g}_{(4)} + \ldots \,.
  }
 Here $\bar{g}_{(2)}$ and $\bar{h}_{(4)}$ are tensors constructed from
$\bar{g}_{(0)}$ using two and four derivatives, respectively.
The expansion breaks down after the logarithmic term, in the sense
that the $\bar{g}_{(n)}$ are no longer covariant tensors.  Fortunately
the first two terms of \gExpand\ are all that we will need.
Explicitly,
  \def\RZero{{\buildrel \scriptscriptstyle{o} \over R}{}}
  \def\aZero{{\buildrel \scriptscriptstyle{o} \over a}{}}
  \eqn{GTwoForm}{
   \bar{g}^{(2)}_{ij} = {1 \over 2} \left( \RZero_{ij} - 
    {1 \over 6} \RZero \, \bar{g}^{(0)}_{ij} \right) \,,
  }
 where $\RZero_{ij}$ is the Ricci tensor of $\bar{g}^{(0)}_{ij}$ and
$\RZero$ is the associated Ricci scalar.

The action under the extremum in \DarkSide\ is
 \eqn{BulkAction}{\eqalign{
   &S_{\rm bulk}[g_{\mu\nu}] + S_{\rm brane}[g_{ij}^{\rm (induced)},\psi] 
      =  \cr
   &\qquad\quad {1 \over 16 \pi G_5} \int_{\cal M} d^5 x \, \sqrt{g}
     \left[ R + 20 \right] +
    {1 \over 16 \pi G_5} \int_{\partial {\cal M}_\epsilon} d^4 \xi \,
     \sqrt{g^{\rm (induced)}} \left[ -2 \Theta + \alpha \right] \,.
  }}
 We have located the cutoff brane on the hypersurface $\partial {\cal
M}_\epsilon$ defined by the equation $\rho = \epsilon$.  We have also
defined
  \eqn{AlphaDef}{
   \alpha = \alpha_0 + {16 \pi G_5 \over \sqrt{g^{\rm (induced)}}}
    {\cal L}_{\rm matter}(g_{ij}^{\rm (induced)},\psi) \,.
  }
 The constant $\alpha_0$ is what we will adjust to balance the tension
of the Planck brane against the bulk cosmological constant.  An
imperfect adjustment would lead to the $dS_4$ induced metric, as
commented on after \EBSimp.  Thus we are not claiming to make headway
on the cosmological constant problem; rather, we are pushing it into
the Planck brane.  The extrinsic curvature term in \BulkAction\ is
necessary in order to have a well-defined variational principle.

Extremizing \BulkAction\ subject to $g_{ij}^{\rm (induced)} = \gamma_{ij}$ 
can be achieved by letting $ds_5^2$ have the form \FeffMetric\ and then
setting $\bar{g}_{ij} = \epsilon \gamma_{ij}$ at $\rho = \epsilon$
(this is at least true up to errors which will be subleading in a
derivative expansion).  Then (cf.~(10) of \cite{hs})
  \eqn{ExtremizedBulk}{
   \mathop{\rm extremum}_{g_{ij}^{\rm (induced)} = \gamma_{ij}} \left(
       S_{\rm bulk}[g_{\mu\nu}] + S_{\rm brane}[g_{ij}^{\rm (induced)},\psi]
       \right) = {1 \over 16 \pi G_5} \int d^4 \xi \, {\cal L}
  }
 where
  \eqn{EllDef}{\eqalign{
   {\cal L} &= 4 \int_\epsilon {d\rho \over \rho^3} \sqrt{|\det\bar{g}|} + 
       \left[ {1 \over \rho^2} (-8 + 4 \rho \partial_\rho + \alpha) 
       \sqrt{\bar{g}} \right]_{\rho = \epsilon}  \cr
    &= \sqrt{|\det\bar{g}_{(0)}|} \left[
      {\alpha - 6 \over \epsilon^2} + 
      {\alpha \over 2\epsilon} \tr \bar{g}_0^{-1} \bar{g}_2 -
       \log \epsilon \, \aZero_{(4)} + \hbox{(finite)} \right] \,.
  }}
 Here we have defined 
  \eqn{aZeroDef}{
   \aZero_{(4)} = -{1 \over 8} \RZero^{ij} \RZero_{ij} + 
    {1 \over 24} \RZero^2 \,.
  } 
 This quantity was identified in \cite{hs} as the conformal anomaly of
the CFT.  The AdS/CFT prescription as detailed there is simply to
remove the terms that diverge as $\epsilon \to 0$ via local
counterterms.  This is the only sensible course if the ultimate goal
is to take $\epsilon \to 0$ so that the cutoff boundary becomes the
true boundary.  Instead we want to keep the cutoff boundary at a
finite, arbitrary $\epsilon$ and regard the induced metric
$\gamma_{ij}$ on $\partial {\cal M}_\epsilon$ as the Einstein metric
of the four-dimensional world.  Rewriting \EllDef\ in terms of
$\gamma_{ij}$, one finds
  \eqn{EllSimp}{
   {\cal L} = \sqrt{|\det\gamma|} \left[
    \alpha - 6 + {1 \over 2} R - \log\epsilon \, a_4 + \ldots \right] \,,
  }
 where now $R$ is the Ricci scalar of the metric $\gamma_{ij}$ and
$a_4$ is defined as in \aZeroDef, only using curvature tensors
pertaining to $\gamma_{ij}$ rather than to $\bar{g}^{(0)}_{ij}$.  One
might fear that the logarithmic term in \gExpand\ would contribute to
the $\log\epsilon$ term in \EllSimp.  It doesn't because $\tr
\bar{g}_{(0)}^{-1} \bar{h}_{(4)} = 0$.

Because powers of $\epsilon$ cancel in \EllSimp\ (and $\epsilon$ is
finite anyway) there is no longer an expansion parameter in \EllSimp.
The expansion can only be justified as a derivative expansion,
provided that the embedding of the cutoff brane in the
five-dimensional Einstein space involves only curvatures which are
slight on the length scale $L$.  Combining \AlphaDef\ and
\ExtremizedBulk\ with \EllSimp, setting $\alpha_0 = 6$, and
repristinating powers of $L$, we find
  \eqn{FinalDark}{\eqalign{
    &\mathop{\rm extremum}_{g_{ij}^{\rm (induced)} = \gamma_{ij}}
      \left( S_{\rm bulk}[g_{\mu\nu}] + 
        S_{\rm brane}[g_{ij}^{\rm (induced)},\psi] \right) =  \cr
    &\qquad\quad{}  
     {L \over 32 \pi G_5} \int d^4 \xi \, \sqrt{\gamma} R +
      \int d^4 \xi \, \sqrt{\gamma} 
       {\cal L}_{\rm matter}(\gamma_{ij},\psi) + 
       W_{\rm CFT}[\gamma_{ij}]
  }}
 where $W_{\rm CFT}$ includes the $\log\epsilon$ term in \EllSimp\
plus all the other terms which we indicated with $\ldots\,$.  We indeed
verify the relation $G_4 = 2 G_5/L$.  Also, since the Ricci scalar
came in with the right sign in \FinalDark, the signs of \EinsteinBold\
are consistent.  The calculation leading to \FinalDark\ is similar to
Kaluza-Klein reduction, the main difference being that the relation
$G_4 = 2 G_5/L$ does not involve the total length of the fifth
dimension (which could be infinite), but rather the curvature scale of
the five-dimensional geometry.  This makes the current scenario rather
different from those of \cite{Dimopoulos}, where the circumference of
the extra dimensions does affect the four-dimensional Planck length.

Extremizing \BulkAction\ with respect to $g_{\mu\nu}$ without requiring
$g_{ij}^{\rm (induced)} = \gamma_{ij}$ would amount, 
at leading order in
derivatives, to setting the right hand side of \EinsteinBold\ to zero, as
well as satisfying the bulk equations of motion.
Given some information regarding the structure of the Planck brane, 
higher derivative corrections to \ThetaReq\ and to the right hand side
of \EinsteinBold\ would be accessible through a more meticulous treatment
of this unrestricted extremization problem.  The trick of
\DarkSide\ is to extremize first with the induced metric held fixed
and then argue that the
extremization that remains to be carried out gives us the equations of
four-dimensional gravity (and brane matter if we want it), 
plus something non-local
which we called $W_{\rm CFT}$.  The claim that this something arises
equivalently by integrating out a CFT below a cutoff $\Lambda$ 
is the substance of AdS/CFT and
the basis for the suggestions in \cite{juanPrivate,HV,WittenComment}.

The argument \FeffMetric-\FinalDark\ stands in relation to the
observation \cite{RSalt} of a normalizable graviton approximately as
the derivation of the low-energy effective action of string theory via
beta-functions stands in relation to the calculation of the massless
string spectrum.

\begin{figure}
   \centerline{\psfig{figure=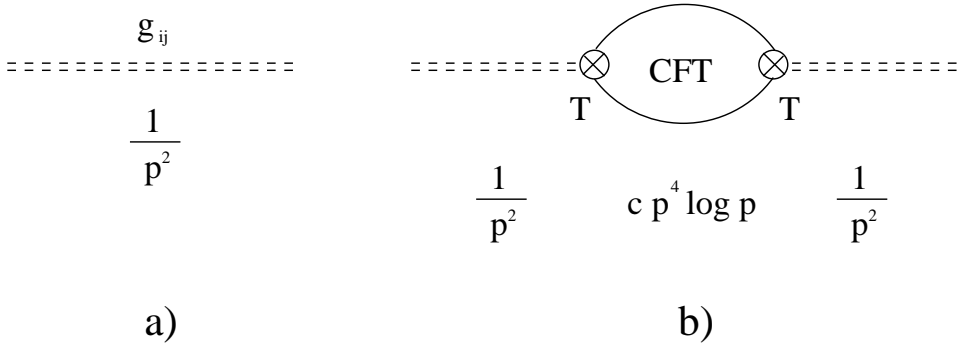,width=5in}}
 \caption{Contributions to the graviton propagator, following
\cite{WittenComment}: a) free graviton propagation; b) leading CFT
correction.  The blob between the stress tensor insertions is intended
to denote the full $\langle TT \rangle_{\rm CFT}$
correlator.}\label{figB}
  \end{figure}

It should be possible to relate corrections to Einstein's equations and hence
Newton's force law directly to the anomaly term $a_4$, proceeding along
the lines of \cite{LiuTseytlin}.  However it is more
transparent to follow the analysis of \cite{WittenComment}, where we
merely differentiate $W_{\rm CFT}[\gamma_{ij}]$ twice with respect to
$\gamma_{ij}$ to obtain the first correction to the graviton
propagator (see figure~\ref{figB}).
 The position space two-point function of the CFT stress tensor has
the form $\langle T(x) T(0) \rangle \sim c/x^8$.  In momentum space
this is $\langle T(p) T(-p) \rangle \sim c \, p^4 \log p$.  The corrected
graviton propagator is
  \eqn{PropCorrect}{\eqalign{
   G^{(2)}(p) &\sim {1 \over p^2} + 
    {1 \over p^2} \ell_{\rm Pl} 
     \left( c \, p^4 \log p \right) \ell_{\rm Pl} {1 \over p^2}  \cr
   G^{(2)}(x) &\sim {1 \over x^2} + 
    {c \, \ell_{\rm Pl}^2 \over x^4} \,,
  }}
 where the factors of the four-dimensional Planck length are vertex
factors for the coupling of the stress tensor to the graviton.  The
altered propagator gives rise to deviations from Newton's $1/r^2$
force law, estimated already from the AdS side in \cite{RSalt}:
  \eqn{ForceCorrect}{
   F = {G m_1 m_2 \over r^2} \left( 1 + 
    a_1 {L^2 \over r^2} + \ldots \right) \,,
  }
 where $a_1$ is a dimensionless number on the order of unity.  
In \ForceCorrect\ I have used \cStandard\
and \GFiveFour\ to combine $c \, \ell_{\rm Pl}^2$ into $\pi L^2/4$.  I do
not claim any originality for the computation in \PropCorrect\ and
\ForceCorrect.  The only further addition I would make to the recorded
comments in \cite{WittenComment} is that the coefficient of leading
correction is indeed computable from the CFT side: up to factors of order
unity it is $G_4$ times the central charge of the CFT.

Clearly, by differentiating \DarkSide\ and keeping track of all the
Lorentz structure we could obtain the corrected propagator in complete
detail and extract the exact value of $a_1$.  I will refrain from
entering into this computation here because another group is pursuing
similar lines \cite{GKR}.  It was important however to present the
general outline of the analysis because it will figure prominently in
the next section.

\section{Bounds and estimates}
\label{Estimate}

Note that \cStandard\ and \GFiveFour\ together imply that the central
charge is $c = {\pi \over 4} {L^2 \over \ell_{\rm Pl}^2}$, where as usual
$\ell_{\rm Pl}$ is the four-dimensional Planck length.  For $AdS_5$
backgrounds arising from type~IIB geometries including D3-branes, $c \sim
N^2$ where $N$ is the number of D3-branes.  So $N \sim L/\ell_{\rm Pl}$.
To be definite, let us suppose that $L$ is on the order of a micron.
Direct measurements of gravity already restrict $L \lsim 1 \, {\rm mm}$,
and proposed experiments might probe Newton's force law to distances as
small as a micron.  $L \sim
1 \, \mu{\rm m}$ means $N \sim 10^{29}$.  This number seems on the
high side for a string compactification: something has to soak up all the
five-form flux.  D3-brane charge is conserved, so it is true that if
we managed to set $N = 10^{29}$ through some arcane string theory
construction, we wouldn't worry about it wiggling.  As disciples of
AdS/CFT we would also be relieved that five-dimensional quantum
gravity effects aren't an immediate problem.  However, a large hidden
CFT is very dangerous in cosmology.  Nucleosynthesis, for example,
would be spoiled if $\rho_{\rm CFT} \gsim \rho_{\rm SM}$, where
$\rho_{\rm SM}$ is the energy density of the Standard Model fields.
Let us assume then that $\rho_{\rm CFT} \ll \rho_{\rm SM}$ around the
time of nucleosynthesis.  Because the CFT has a large number of
degrees of freedom as compared to the Standard Model, this is possible
only if the CFT is much colder than Standard Model excitations.  Suppose 
that the Standard
Model and the CFT are to a good approximation decoupled.  Then
$\rho_{\rm CFT}$ and $\rho_{\rm SM}$ decrease in fixed ratio during
the radiation-dominated era, up to factors of order unity associated
with freezing out the various massive fields of the Standard Model.
In the matter-dominated era, $\rho_{\rm CFT}$ and $\rho_{CBR}$ decrease
in fixed ratio.  So we can guarantee that nucleosynthesis is unaffected by
the CFT if we demand $\rho_{\rm CFT} \ll
\rho_{\rm CBR}$ today.  This translates roughly to $T_{\rm CFT} \lsim
T_{\rm CBR}/c^{1/4} \approx 10^{-14} \, {\rm K}$ today if we want $L$
on the order of a micron.\footnote{We have used the AdS/CFT prediction
$\rho \sim cT^4$.  Naively counting flat directions in ${\cal N}=4$
super-Yang-Mills theory suggests $\rho \sim \sqrt{c} T^4$.  Even if
this were somehow true for a special CFT, it would only soften
\eno{NScaling} to $T_{\rm CFT} \lsim T_{\rm CBR}/N^{1/4}$.}  To
summarize,
  \eqn{NScaling}{
   c \sim N^2 \,, \qquad\ L \sim N \ell_{\rm Pl} \,, \qquad\ 
    T_{\rm CFT} \lsim T_{\rm CBR}/\sqrt{N} \,.
  }

\begin{figure}
   \centerline{\psfig{figure=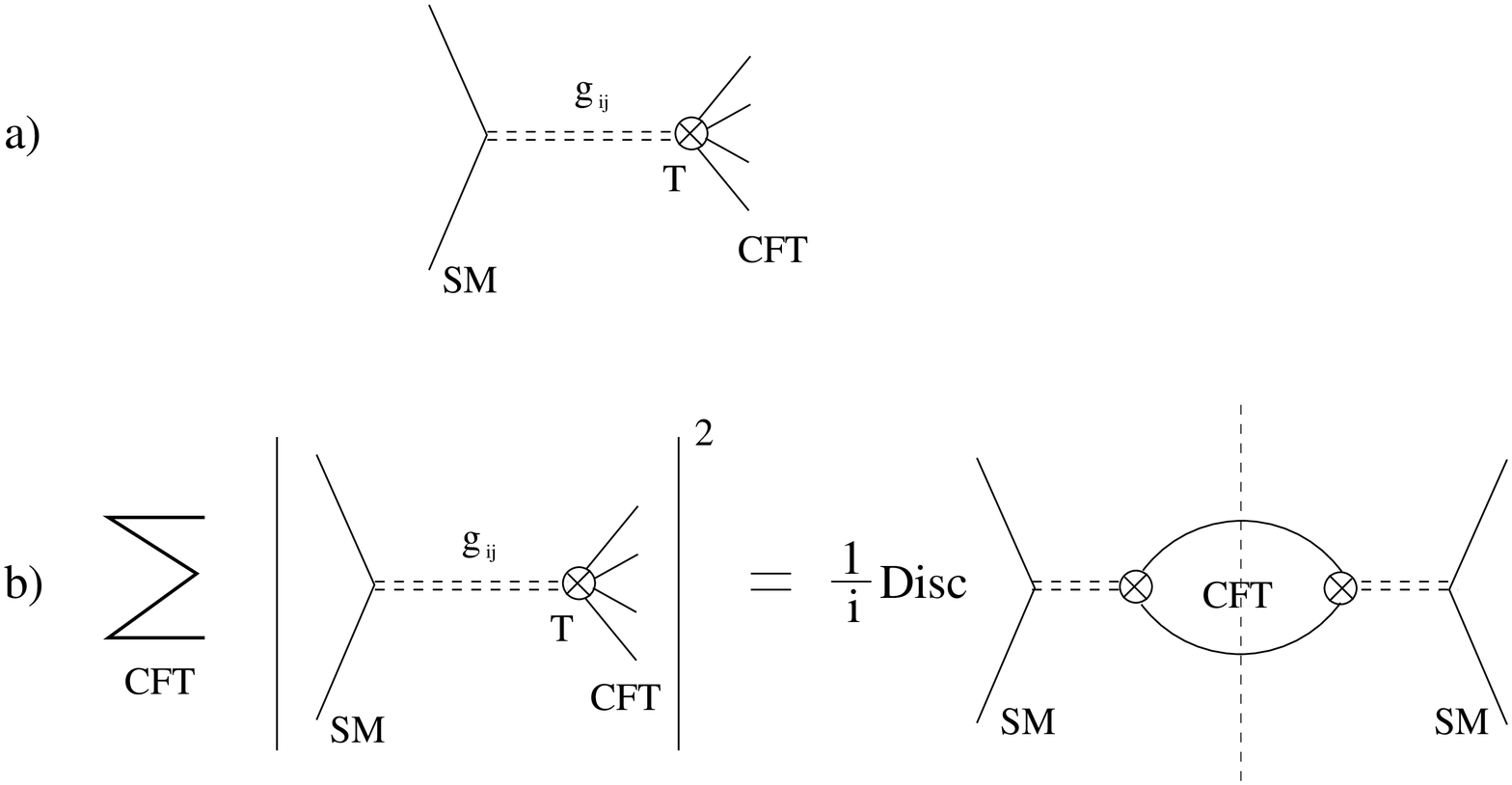,width=6in}}
 \caption{a) Standard Model particles losing energy via graviton
exchange to CFT excitations.  b) The inclusive rate is given by a
unitarity cut of the first correction to the graviton
propagator.}\label{figA}
  \end{figure}

Suppose the CFT is cold enough at some early time to satisfy $\rho_{\rm
CFT} \ll \rho_{\rm SM}$.\footnote{Section~\ref{FRW} treated the opposite
limit.  One should be able to use the equation \EinsteinBold\ with $T^{\rm
(CFT)}_{ij}=0$ to find a hypersurface in $AdS_5$ whose induced metric is
real-world cosmology.  But this is only an equivalent means to find what we
already know by solving Einstein's equations.  In this section we will
``cast down the ladder'' and work directly in four dimensions whenever
possible.}  Cosmology at later times could still be spoiled if energy leaks
too quickly from visible matter into the CFT.  The analogous problem in
theories with compact extra dimensions is cooling by emission of bulk
gravitons \cite{DimopoulosTwo}.  To evaluate whether there is a problem in
our case, we must investigate the mechanisms of thermal equilibration
between the CFT and the other matter in the universe, operating on the
assumption that the CFT is very cold.  Fortunately the tools are already
partly in hand.  Standard Model particles can lose energy to the conformal
field theory through processes controlled by the graph in
figure~\ref{figA}a).  The inclusive rates from these graphs are related to
the $1/r^4$ correction to Newton's law through the unitarity relation
illustrated in figure~\ref{figA}b).
 In particular, the inclusive rate goes as $\ell_{\rm Pl}^2 L^2$.  By
dimensional analysis the contribution they make to the loss of
Standard Model energy density over time is 
  \eqn{RhoLoss}{
   \left( {d\rho \over d\tau} \right)_{\rm lost} = -a_2 
    \ell_{\rm Pl}^2 L^2 T_{\rm SM}^9 \,,
  }
 where $a_2$ is a dimensionless number of order unity and $T_{\rm SM}$
is the temperature of Standard Model excitations.  From an AdS point
of view, \RhoLoss\ is literally the rate at which energy density falls
across the horizon to be absorbed by the D3-branes.  Three powers of
$T_{\rm SM}$ come from the absorption cross section
\cite{IgorAbsorb,gktAbsorb,gkSchwing}; also there are powers of
$T_{\rm SM}$ from the finite temperature kinematics of the Standard
Model particles.  Energy density also decreases because of Hubble
expansion: in total,
  \eqn{RhoEvolve}{
   {d\rho_{\rm SM} \over d\tau} = -{\dot{a} \over a} T_{\rm SM}^4 - 
    \ell_{\rm Pl}^2 L^2 T_{\rm SM}^9 \,,
  }
 where we have dropped factors of order unity.  One such factor is the
central charge of the Standard Model fields which are light compared to the
temperature at any given time.  Approximately this same factor appears in
both terms on the right hand side of \RhoEvolve, so it doesn't matter much
for the relative size of the terms.  However $a_2$ does matter, and it
should be computed if a more accurate estimate than the one presented here
is desired.  We have also suppressed a term in \RhoEvolve\ for CFT energy
leaking back into visible fields, but that is OK since we are operating on
the assumption that the CFT is cold.

To determine whether the CFT is appreciably affecting the cosmology,
one should compare the two terms in \RhoEvolve.  Their ratio is
  \eqn{FMerit}{
   \varkappa = \ell_{\rm Pl}^2 L^2 T_{\rm SM}^5 H^{-1}
  }
 where $H^{-1} = a/\dot{a}$ is the inverse Hubble time (a function of
$\tau$).  The CFT will not appreciably affect cosmology as long as
$\rho_{\rm CFT} \ll \rho_{\rm SM}$ and $\varkappa$ is small.  What
small means in this context depends on all the ``factors of order
unity'' that we have dropped.  All these factors are calculable: once
we have \cStandard\ and \GFiveFour\ the rest is essentially
kinematics.  In order to make some preliminary estimates I will assume
that the Hubble expansion term in \RhoEvolve\ dominates over the
energy loss term when $\varkappa \ll 1$.

The state of the universe today does not lead to a dramatic bound on
$L$: for instance, estimating $\kappa$ for the rate of energy loss from the
CBR to the CFT gives
  \eqn{LBoundToday}{
   L^2 \ll {1 \over \ell_{\rm Pl} H_o^{-1} T_{\rm CBR}^5} \sim 
    10^{33} \, {\rm cm}^2 \,,
  }
 which is easily passed by any realistic theory.  However the bound
tightens as one goes back in time.  Tracing the matter-dominated
cosmology back to the time of last scatter at $z \sim 10^4$, one
obtains roughly
  \eqn{LBoundLastScatter}{
   L^2 \ll 10^{33} \, {\rm cm}^2 
    \left( {a_{\rm last\ scatter} \over a_o} \right)^5 
    {H_o^{-1} \over H_{\rm last\ scatter}^{-1}}
    = 10^{33} \, {\rm cm}^2 
       \left( {a_{\rm last\ scatter} \over a_o} \right)^{7/2}
    = 10^{19} \, {\rm cm}^2 \,,
  }
 still not meaningfully restrictive.  Tracing the radiation-dominated
cosmology back to nucleosynthesis at $z \sim 10^{10}$, one obtains 
  \eqn{LBoundNucleosynthesis}{
   L^2 \ll 10^{19} \, {\rm cm}^2 
    \left( {a_{\rm nucleosynthesis} \over a_{\rm last\ scatter}} 
     \right)^5 
     {H_{\rm last\ scatter}^{-1} \over H_{\rm nucleosynthesis}^{-1}}
    = 10^{21} \, {\rm cm}^2 
       \left( {a_{\rm nucleosynthesis} \over a_{\rm last\ scatter}} 
        \right)^3
    = 10 \, {\rm cm}^2 \,.
  }
 Still this bound is satisfied with four orders of magnitude to spare
(in $L$) if we suppose $L$ to be on the order of a micron.  I
emphasize the extreme simple-mindedness of the estimates: all I have
done in \LBoundNucleosynthesis\ is to write
  \eqn{ReallySimple}{
   L^2 = \varkappa \, {1 \over \ell_{\rm Pl}^2 T_{\rm SM}^5 H^{-1}}
       \approx \varkappa \, {1 \over 
        \ell_{\rm Pl}^2 T_{\rm CBR}^5 H_o^{-1}}
       {1 \over z_{\rm last\ scatter}^{7/2}} \,
       \left( z_{\rm last\ scatter} \over z_{\rm nucleosynthesis} 
        \right)^3 \,,
  }
 and then demand $\varkappa \ll 1$.  The powers of $z$ in
\LBoundLastScatter-\ReallySimple\ arise from the relations $H^{-1} \sim
a^{3/2}$ for the matter dominated cosmology and $H^{-1} \sim a^2$ for
the radiation-dominated cosmology.  In view of the actual number
obtained in \LBoundNucleosynthesis, a more accurate estimate would be
desirable.  One can also attempt to trace cosmology back to larger $z$
and tighten the bound on $L$ further, if one feels convinced that
$\varkappa$ must still be small for $z > 10^{10}$.

An independent bound on $L$ could obtained by checking the effect on
supernovas of energy loss to the CFT, as in \cite{DimopoulosTwo}.  The
energy scales here are on the order of $30 \, {\rm MeV}$, so a
slightly better bound than \LBoundNucleosynthesis\ might be expected.

There is yet another way to set a bound on $L$ if we assume that the
$AdS_5$ geometry comes from type~IIB string theory through some
Freund-Rubin ansatz or related compactification.  In such
compactifications, the extra five dimensions have the same length
scale $L$ as $AdS_5$.  Suppose $AdS_5 \times S^5$ is the relevant
geometry.  Then the standard string theory relation $16 \pi G_{10} =
(2\pi)^7 g_s^2 \alpha'^4$ combined with \GFiveFour\ and $\Vol S^5 =
\pi^3 L^5$ leads us to
  \eqn{LGalpha}{
   L = g_s^{1/3} \left( 16 \pi^3 \alpha'^4 \over G_4 
    \right)^{1/6} \,.
  }
 Type~IIB theory has an S-duality symmetry which takes $g_s \to
1/g_s$.  Thus we can assume that $g_s \leq 1$.  A conventional value
of $\sqrt{\alpha'}$ would be only a few times the four-dimensional
Planck length, $\ell_{\rm Pl}$.  This results in a bound on $L$ which
is also a few times $\ell_{\rm Pl}$.  In order to make $L$ observably
big, we would have to make $\sqrt{\alpha'}$ big too.  What is the
biggest $\sqrt{\alpha'}$ we could possibly imagine?  In the old days
of string theory the answer would have been $\sqrt{\alpha'} \sim 1 \,
{\rm GeV}^{-1} \approx 0.2 \, {\rm fm}$: this is literally the Regge
slope of observed hadronic spectra.  In recent literature
\cite{Antoniadis}, values of $\sqrt{\alpha'}$ as big as $1 {\rm
TeV}^{-1}$ have been regarded as acceptable.  Plugging these numbers
into \LGalpha\ leads to 
  \eqn{ReggeBound}{\eqalign{
   L \lsim 10^{-7} \, {\rm cm} &\qquad
    \hbox{for $\sqrt{\alpha'} \sim 1 {\rm GeV}^{-1}$}  \cr
   L \lsim 10^{-11} \, {\rm cm} &\qquad
    \hbox{for $\sqrt{\alpha'} \sim 1 {\rm TeV}^{-1}$} \,.
  }}
 The bad news is that deviations from Newton's force law on length
scales this small won't be detected any time soon.  The good news is
that standard cosmology is no problem, as far back as nucleosynthesis
and further.  If we assume that the radiation-dominated FRW solution
still pertains, we can estimate the redshift $z_*$ and the thermal
energies $T_*$ at which $\varkappa = 1$.  The result is
  \eqn{SourEnergies}{\eqalign{
   z_* \sim 10^{15} \,, \quad T_* \sim 100 \, {\rm GeV} 
     &\qquad \hbox{for $\sqrt{\alpha'} \sim 1 {\rm GeV}^{-1}$}  \cr
   z_* \sim 10^{18} \,, \quad T_* \sim 100 \, {\rm TeV}
     &\qquad
    \hbox{for $\sqrt{\alpha'} \sim 1 {\rm TeV}^{-1}$} \,.
  }}
 To get the numbers in \SourEnergies\ we have combined several
approximations and assumptions.  The ``error in the exponent'' should
probably be taken to be about $\pm 2$.  It is somewhat suggestive that
the values of $T_*$ we found are ``within errors'' of the boundary of
our direct knowledge of particle physics.  If the string scale is at a
${\rm TeV}$, then physics changes sufficiently there that we can no
longer have any confidence that the radiation-dominated FRW cosmology
is relevant.  Thus the second line of \SourEnergies\ only shows that
there are no cosmological problems as far back as we can trace the
theory.  Strings at a ${\rm GeV}$ are a different matter, and we will
return to them shortly.

Although type~IIB string theory provides the best-understood vacua
involving $AdS_5$, it is conceivable that some other type of string
theory, even a non-critical string, could have an $AdS_5$ vacuum: see
for example \cite{PolWall}.  For a non-critical string, \LGalpha\
would not be the right estimate, since some or all of the five compact
dimensions simply aren't there.  Suppose the non-critical string lives
in $n$ dimensions, with $n \geq 5$.  Assume also that it exhibits some
form of S-duality, so that the coupling cannot be parametrically
large.  Then
  \eqn{NonCritical}{
   L \lsim \left( \sqrt{\alpha'} \over 
    \ell_{\rm Pl} \right)^{\gamma} \sqrt{\alpha'}
  }
 where $\gamma = 2/(n-4)$.  If we allow $n$ to range from $10$ to $5$,
the corresponding range of $\gamma$ is from $1/3$ to $2$.  It is also
conceivable that some intersecting configuration of branes in critical
string theory could have an $AdS_5$ component in its near-horizon
geometry, and a different relation from \LGalpha\ could pertain if
some of the branes had more than $3+1$ worldvolume dimensions.  I am
not currently aware of any completely well-defined, non-critical string theory
other than the $c \leq 1$ toys.  Nor can I give a string theoretic example of
intersecting branes with an $AdS_5$ near-horizon geometry.  Besides,
if the extra dimensions of the intersecting branes are larger than
$L$, then the salient physics of extra dimensions would be more along
the lines of \cite{Dimopoulos} than \cite{RSalt}.  For the sake of a
concrete discussion, let us stick to \LGalpha, with \NonCritical\ as
a possible alternative.

Once we have ventured to set $\sqrt{\alpha'} \approx 1 \, {\rm GeV}^{-1}$,
the burning question is why all collider physics from a ${\rm GeV}$ up to a
${\rm TeV}$ isn't dramatically different.  The simplest answer is rather
iconoclastic.  It is that from a four-dimensional point of view, strings
are nothing more than QCD flux tubes.  For energies well above $1 \, {\rm
GeV}$, but below the cutoff scale $\Lambda \sim L/\alpha'$, a better set of
variables is the particles of the Standard Model, plus a massless
propagating graviton.  In the low-energy regime where strings are the good
variables, there is a massless graviton in the closed string spectrum.  The
graviton must be present in a description of the theory at any scale: on
very general grounds \cite{WeinbergWitten} it is impossible for the graviton
to be a composite particle.\footnote{I thank M.~Strassler and R.~Sundrum
for discussions on this point and related issues.}  The spectrum could also
include massless open strings if the Planck brane involves D-branes.  The
gluons in the Standard Model lagrangian might be represented in this way at
low energies.  Intuitively, the reason why a disk diagram with two gluon
boundary insertions and one bulk insertion of a graviton wouldn't couple
gluons to gravity on the scale of femtometers is that the wavefunction
overlap is small.  This is the magic of extra dimensions (exploited
similarly in \cite{Dimopoulos}): \LGalpha\ is roughly a condition on how
big the extra dimensions have to be in the well-understood type~IIB string
theory examples to make low-energy strings consistent with gravity at the
four-dimensional Planck scale.  The modified relation \NonCritical\ could
be pertinent for alternative models, as discussed in the previous
paragraph.

The view taken in the previous paragraph is distinct from those of
\cite{witHolTwo} or \cite{HV}.  Wilson loops in AdS/CFT usually seek
out a location of large redshift in the bulk geometry in order to
lower their tension to the scale of confinement.  The current scenario
has Wilson loops terminating on the Planck brane, and the relevant
geometry is the geometry near the Planck brane.  I would not exclude
scenarios where a large redshift does exist near the Planck brane, and
the parameter entering into the Regge relation is a redshifted
$\alpha'$.  If that is the way we think QCD strings are realized, then
once again the bound on $L$ is tighter than $L \lsim 1 \, {\rm nm}$,
since the $\alpha'$ that enters \LGalpha\ is the 
un-redshifted string tension.

Strings at a ${\rm GeV}$ seem like a natural apotheosis of the proposals of
\cite{Dimopoulos,Antoniadis}.  We do not have to ``get rid'' of the
graviton if there are extra dimensions on the scale of a nanometer.
(Significantly smaller $L$ would work in a model where \NonCritical\
pertains).  We do not have to worry about nucleosynthesis if the estimate
\LBoundNucleosynthesis\ bears out.  But we do have to face some hard
questions.  First, if $L \sim 1 \, {\rm nm}$, how do we manage to
accommodate $N = L/\ell_{\rm Pl} \approx 10^{26}$ D3-branes?  Something has
to soak up all the Ramond-Ramond flux, and that sounds like an impossible
stretch for string compactifications (see for example \cite{HV}).  $N$
comes out somewhat smaller if $g_s$ is small, or in models where
\NonCritical\ pertains with $n<10$.  Second, string theory would have to
face up to hadron physics in the energy range between pions and partons.
Regge trajectories are as suggestive as they always were, but there is much
more to be explained.  Processes where some or all of the final energy
winds up in CFT excitations are likely to be a problem.  However the
relevant branching ratios typically depend on $L$ rather than $\alpha'$,
and amount to yet another way of setting an upper bound on $L$.  Third,
strings could stretch from the Planck brane all the way into the $AdS_5$
bulk (to connect with a D3-brane if one wants to think in those terms) at
only a finite cost in energy.  The mass of such a string is roughly
$L/\alpha'$, which comes out to be approximately $3000 \, {\rm TeV}$ if we
use $L \sim 1 \, {\rm nm}$.  This is out of the range of colliders, but it
is nevertheless a dangerous number for any sort of loop computation because
these strings are so numerous: there are as many of them as there are
D3-branes.  Their mass is bigger if \NonCritical\ applies: 
$L/\alpha' \sim 10^{18 \gamma} \, {\rm GeV}$.  Fourth and finally, if flux
tubes are long strings ending on the Planck brane, then what are quarks?

As observed in section~\ref{General}, 
there is a precise way of characterizing the strings stretched from the
Planck brane to the horizon of $AdS_5$: they are the massive $W$ bosons
associated with the separation of the Planck brane from the $L/\ell_{\rm
Pl}$ D3-branes that create the $AdS_5$ geometry.  The scale of these masses 
could be lowered, say to $30 \, {\rm TeV}$, if $L$ falls
sufficiently short of saturating the bound in the first line of
\ReggeBound.  Or we could return to strings at a ${\rm TeV}$ and get
approximately the same $30 \, {\rm TeV}$ Higgs scale by saturating the
bound in the second line of \ReggeBound.  Either way, we are left with a
version of \cite{VafaFrampton}, only with an enormous hidden sector gauge
group and strings at a ${\rm GeV}$ or a ${\rm TeV}$.  In
\cite{VafaFrampton}, it seemed like coupling the CFT to gravity might
resurrect the hierarchy problem.  This is less of a problem if the string
scale is smaller than or comparable to the scale of soft breakings of the
CFT: one may hope that stringy ``softness'' ameliorates the divergences of
gravity already at the string scale.  There is no clear microscopic picture
of what the theory is without specifying the nature of the Planck brane.
However, the relation \LGalpha\ between Newton's coupling and other 
low-energy quantities should not depend on the detailed properties of
the Planck brane.

In conclusion, insisting that $AdS_5$ has to come from string theory
provides a bound on $L$ which is sharper than we were able to obtain
from nucleosynthesis, and which appears to rule out experimental
observation of \ForceCorrect.  There are two reasons why string theory
demands a small $L$.  First, $L/\ell_{\rm Pl} \sim N$, where $N$ is
the number of units of Ramond-Ramond five-form flux.  It is hard to
make this number really big in string compactifications.  Second, $G_4
L^6 \lsim \alpha'^4$, so we can only get big $L$ if we allow big
$\alpha'$.  In an attempt to be maximally optimistic about the size of
$L$, we have reconsidered strings at a ${\rm GeV}$.  Even this radical
step only gave us $L \lsim 1 \, {\rm nm}$.  If we make $\alpha'$ even
bigger, it only heightens the difficulties we encountered trying
to make sense of ${\rm GeV}$ strings.  The strategies proposed in
AdS/CFT contexts to relate strings to QCD flux tubes generally have
the property that the fundamental $\alpha'$ is smaller than $1 \, {\rm
GeV}^{-2}$, implying a tighter bound on $L$.

\section{Discussion}
\label{Discussion}

The FRW cosmology found in section~\ref{FRW} is an interesting check
of the claim that the ``alternative to compactification'' proposed in
\cite{RSalt} is equivalent to a cutoff conformal field theory coupled to
four-dimensional gravity.  However, as emphasized in
section~\ref{Estimate}, the CFT shouldn't make any sizeable
contribution to the actual cosmology of our universe at times later
than $z = 10^{10}$.  Before that time, one is entitled to speculate
about the physical relevance of the solution of section~\ref{FRW}.
Suppose that the CFT and the visible sector matter on the Planck brane
were in thermal equilibrium at some early time.  Assuming that the CFT
has a much larger central charge, we have $\rho_{\rm CFT} \gg
\rho_{\rm matter}$, and the solution found in section~\ref{FRW} should
approximately describe the cosmology.  At late times one needs
$\rho_{\rm CFT} \ll \rho_{\rm matter}$.  In parallel with
\cite{DimopoulosTwo}, we might imagine an inflationary scenario where
the inflaton lives on the Planck brane.  Then reheating directly
affects only the visible sector, and if $\varkappa$ is small by the
time of reheating there is substantially no thermal equilibration with
the CFT.

In a scenario with ${\rm GeV}$ strings, thermalization with the CFT
sets in significantly around $100 \, {\rm GeV}$ (although we must
recall that the estimates here were extremely crude).  That alone
might lead us to rule this case out unless a reheating mechanism could
be proposed at a lower scale.

In known string compactifications, the number $N$ of D3-branes is
typically on the order of $10$.  As many as $10^3$ D3-branes were
claimed to be attainable in certain orbifold examples \cite{HV}.  If
we take this as a strict bound, then the relation $L/\ell_{\rm Pl}
\sim N$ puts our entire discussion at an inaccessibly small length
scale: $L \sim 10^{-30} \, {\rm cm}$ for $N = 10^3$.  (As usual,
$\ell_{\rm Pl}$ is the four-dimensional Planck scale).  The formalism
developed in section~\ref{General} could still be useful for
extracting a ``low-energy'' effective theory---``low-energy'' being
interpreted now as much less than $10^{16} \, {\rm GeV}$.  Standard
inflation occurs around $10^{14} \, {\rm GeV}$, so it is possible one
might embed a ``conventional'' inflationary model in $AdS_5$ using the
$dS_4$ solution discussed after \EBSimp.  The amusing aspect of such a
model is that there is a natural candidate for the pre-inflationary
universe: it is the radiation-dominated FRW solution found in
section~\ref{FRW}.

There are two solid conclusions to be drawn from the estimates of
section~\ref{Estimate}.  First, our present understanding of
nucleosynthesis would not be threatened if deviations from Newton's
force law of the form \ForceCorrect\ were found.  We already know that
such deviations cannot be present on scales much larger than a
millimeter, and this is enough to suppress the associated loss of
energy to the conformal field theory for $z$ as large as $10^{10}$.
Second, string theory as we understand it seems to forbid an $AdS_5$
space large enough to cause measurable deviations from Newton's force
law.  Even if we are willing to take the string scale down to $1 \,
{\rm GeV}$ and regard strings as collective effects of QCD, $L$ still
can't be larger than $1 \, {\rm nm}$.

There is nothing sacred about an $AdS_5$ bulk spacetime: it has been
the focus of so much recent literature in part because it is simple.
Practically any string theory realization of $AdS_5$ will include
scalar fields, and if they have a non-trivial profile, large
deviations from $AdS_5$ are the generic behavior far from the 
boundary.  The literature on
renormalization group in AdS/CFT flows provides ample evidence of this
(see for example
\cite{GPPZfirst,DZ,gDil,kSfetsos,fgpwOne,GPPZone}).
Only a subset of these geometries can support
finite temperature, due to boundary conditions on the scalars at the
black hole horizon.  A felicitous feature of $AdS_5$, which will not
be shared by generic ``RG flow'' geometries, is that the relation $G_4
= 2 G_5/L$ obtains no matter where the Planck brane is in the bulk
geometry.  The formalism worked out in section~\ref{General} will
still retain its general features in a more generic bulk geometry, but
details will be rather different: for instance, it is no longer clear
that the induced metric on the cutoff brane will be the Einstein frame
metric.

A cutoff brane in a bulk geometry whose AdS/CFT dual is a quantum
field theory undergoing renormalization group flow corresponds to
gravity coupled to that same QFT.  The proposal of \cite{RShierarchy}
is to put the Standard Model not on the cutoff brane, but rather on
some brane far from the boundary, where $g_{tt}$ is very small.  There
are several ways that such a construction could be realized in string
theory.  First, if the five-dimensional space-time ends at a finite
minimum of $g_{tt}$, then one can show that the end-of-the-world brane
must have negative tension.  The best-understood constructions in the
current literature which admit negative tension end-of-the-world
branes are type~I$'$ string theory and certain Calabi-Yau
compactifications of Horava-Witten theory.  At the classical level,
these constructions do not allow an $AdS_5$ bulk: there is always some
scalar that evolves across the five-dimensional bulk.  There is not
yet compelling evidence that all scalars could be held fixed and an
$AdS_5$ bulk obtained.  It would be possible to develop a formalism
similar to the one in section~\ref{General} for type~I$'$ or
Horava-Witten constructions, but it would have more the flavor of an
ordinary Kaluza-Klein reduction, where heavy fields are integrated out
and light fields are kept.  The distinctive feature of \DarkSide\ is
that it enables us to obtain a non-local functional which summarizes
the dynamics of {\it infrared} degrees of freedom.

If there are no negative tension branes, the only option is for the
five-dimensional space-time to continue all the way to $g_{tt}=0$.  If
there are scalars involved, the generic behavior is for curvatures to
become strong as $g_{tt} \to 0$.  AdS/CFT has limited computational
power in such circumstances.  The best hope is that string theory
provides a resolution of the strong curvatures.  If visible sector
fields live on branes at strong curvatures, then we are not in a
position to say much about the physics.  It is also conceivable
\cite{RandallLykken} that visible sector fields live on a probe brane
at small but nonzero $g_{tt}$.  There are potential phenomenological
virtues to such a model, but it seems somewhat contrived.

String theory and string dualities have taught us that extra
dimensions are theoretically inexpensive.  But the view of the fifth
dimension espoused in the current paper is not excessively literal:
rather than making the claim that there is actually a large extra
dimension of space waiting to be discovered, the statement is that an
extra dimension is a convenient way to describe collective phenomena
of a strongly coupled quantum field theory---in the present case, a
conformal field theory coupled to gravity.  To make this seem more
definite, suppose measurements of gravity at a micron did after all
turn up deviations from Newton's law of the form \ForceCorrect.  The
``AdS'' interpretation would be that gravitons are propagating in the
fifth dimension, while the ``CFT'' interpretation would be that a loop
of gauge bosons in a purely four-dimensional theory had contributed.
Which interpretation we prefer is a matter of ontology: if AdS/CFT is
right then they are absolutely indistinguishable on experimental
grounds.  My current ontology isn't very happy either
with a CFT with $c \sim 10^{58}$ or with a fifth dimension with
curvatures on the scale of a micron.  But it is in the subtle guises
of string duality and string compactification that I suspect extra
dimensions have the best chance of improving our understanding of the
physical world.

\section*{Acknowledgements}

Although it is tricky to assign credit in the absence of publications, I
should acknowledge the important contributions of J.~Maldacena and
E.~Witten.  As far as I can ascertain, J.~Maldacena was the first to
enunciate the view that the Randall-Sundrum ``alternative to
compactification'' is nothing more nor less than gravity coupled to a
strongly interacting CFT; and as far as my personal knowledge extends,
E.~Witten was the first to suggest a definite calculation based on the
idea, namely a correction to Newton's law based on the two-point function
of the stress tensor of the CFT.  I have also profited greatly from
conversations with H.~Verlinde.  I thank the participants of the conference
``New Dimensions in String theory and Field theory'' at the ITP in Santa
Barbara for lively discussions---especially I.~Klebanov, M.~Gremm,
R.~Myers, V.~Periwal, B.~Ovrut, K.~\hbox{Skenderis}, S.~Giddings, S.~Elitzur,
E.~Witten, V.~Hubeny, L.~Bildsten, R.~Maimon, R.~Sundrum, M.~Strassler, 
and G.~Horowitz.
I thank N.~Warner, K.~Pilch, D.~Freedman, O.~DeWolfe, A.~Karch,
E.~Silverstein, and S.~Kachru for earlier discussions, and particularly
D.~Gross for reading an early version of the manuscript and for useful
comments.

This research was supported by the Harvard Society of Fellows, and
also in part by the NSF under grant number PHY-98-02709, and by DOE
grant DE-FGO2-91ER40654.  I thank the ITP at Santa Barbara for
hospitality while the work was carried out.


\bibliography{frw}
\bibliographystyle{ssg}

\end{document}

%% file: lmacros.tex
\input umacros
\def\lbldef#1#2{\expandafter\gdef\csname #1\endcsname {#2}}
\def\eqn#1#2{\lbldef{#1}{(\ref{#1})}%
\begin{equation} #2 \label{#1} \end{equation}}
\def\eqalign#1{\vcenter{\openup1\jot
    \halign{\strut\span\TL & \span\TR\cr #1 \cr
   }}}
\def\eno#1{(\ref{#1})}

%% file: umacros.tex
\def\TL{\hfil$\displaystyle{##}$}
\def\TR{$\displaystyle{{}##}$\hfil}

\def\TT{\hbox{##}}
\def\seqalign#1#2{\vcenter{\openup1\jot
  \halign{\strut #1\cr #2 \cr}}}

\def\comment#1{}
\def\fixit#1{}

\def\mop#1{\mathop{\rm #1}\nolimits}

\def\Vol{\mop{Vol}}

\def\tr{\mop{tr}}

\def\overleftrightarrow#1{\vbox{\ialign{##\crcr
     $\leftrightarrow$\crcr\noalign{\kern-0pt\nointerlineskip}
     $\hfil\displaystyle{#1}\hfil$\crcr}}}

\def\lsim{\mathrel{\mathstrut\smash{\ooalign{\raise2.5pt\hbox{$<$}\cr\lower2.5pt\hbox{$\sim$}}}}}
\def\gsim{\mathrel{\mathstrut\smash{\ooalign{\raise2.5pt\hbox{$>$}\cr\lower2.5pt\hbox{$\sim$}}}}}

\def\sqr#1#2{{\vcenter{\vbox{\hrule height.#2pt
         \hbox{\vrule width.#2pt height#1pt \kern#1pt
            \vrule width.#2pt}
         \hrule height.#2pt}}}}

\def\href#1#2{#2}  
